\documentclass[aps,prf,twocolumn,reprint]{revtex4}
\usepackage{natbib}	

\usepackage[dvips]{graphicx}
\usepackage[export]{adjustbox}
\usepackage{subfigure}
\usepackage{amsmath}
\usepackage{amssymb}
\usepackage{wasysym}
\usepackage{color}

\begin{document}

\title{Dissipative Effects on Inertial-Range Statistics at High Reynolds numbers}

\author{Michael Sinhuber}
\affiliation{Department of Civil and Environmental Engineering, Stanford University, Stanford, CA 94305, USA}
\author{Gregory P. Bewley\footnote{gpb1$@$cornell.edu}}
\affiliation{Department of Mechanical and Aerospace Engineering, Cornell University, Ithaca, NY 14853, USA}
\author{Eberhard Bodenschatz}
\affiliation{Max Planck Institute for Dynamics and Self-Organization, 37077 G\"{o}ttingen, Germany \\
	Institute for Nonlinear Dynamics, University of Goettingen, 37077 G\"{o}ttingen, Germany \\
	Department of Physics, Cornell University, Ithaca, NY 14853, USA\\
	Department of Mechanical and Aerospace Engineering, Cornell University, Ithaca, NY 14853, USA}

\date{\today}

\begin{abstract}

Using the unique capabilities of the Variable Density Turbulence Tunnel  at the Max Planck
Institute for Dynamics and Self-Organization, G\"{o}ttingen, we report experimental result on classical grid turbulence that uncover fine, yet important details of the structure functions in the inertial range. This was made possible by measuring extremely long time series of up to $10^{10}$ samples of the turbulent fluctuating velocity, which corresponds to $\mathcal{O}\left(10^5\right)$ large eddy turnover times. These classical grid measurements were conducted in a well-controlled environment at a wide range of high Reynolds numbers from $R_\lambda=110$ up to $R_\lambda=1600$, using both traditional hot-wire probes as well as NSTAP probes developed at Princeton University. We found that deviations from ideal scaling are anchored to the small scales and that dissipation influences the inertial-range statistics at scales larger than the near-dissipation range.

\end{abstract}

\maketitle

%\section{Introduction}

One of the distinguishing features of turbulent flows is the deviation of its statistics from Gaussian, resulting in the frequent occurrence of extreme events. Despite decades of research, an exact prediction or description  of the statistics of these extreme events based upon the governing Navier-Stokes equations by \citet{Navier1827} and \citet{Stokes1845} is still absent. The rate of extreme events, such as strong wind gusts in natural turbulent flows are connected to the tails of the probability density function of the longitudinal velocity increments $f\left(\delta u,r\right)$. The moments of this statistical object are the longitudinal structure functions of n\textsuperscript{th} order $S_n = \langle\left(u\left(x\right)-u\left(x+r\right)\right)^n\rangle$. Here, $u\left(x\right)$ is the longitudinal velocity component aligned with the separation $r$ and $x$ the position. One of the few exact results that can be derived from the Navier-Stokes equations under the assumptions of stationarity, homogeneity and isotropy concerns the third-order longitudinal structure function,

\begin{align}
S_3\left(r\right) - 6\nu\frac{\mathrm{d}}{\mathrm{d}r}S_2\left(r\right) = -\frac{4}{5}\langle\epsilon\rangle r+q\left(r\right).
\label{eq:45full}
\end{align}

This relation was derived originally by \citet{Karman1938} for correlation functions and reformulated by \citet{Kolmogorov1941a} in terms of structure functions. Here, $q\left(r\right)$ is a source term containing the information about energy injection, $\nu$ the kinematic viscosity and $\langle\varepsilon\rangle$ the mean energy dissipation rate. In the limit of infinite Reynolds numbers, $\nu\rightarrow0$, the second term on the left-hand side of the equation vanishes as long as the derivative remains finite. Following the classical cascade picture by \citet{Richardson1922} and the original arguments by \citet{Kolmogorov1941b} , there is an intermediate range of scales where neither the energy injection at the large scales nor energy dissipation at small scales influence the statistics of the flow. In the inertial range, one obtains Kolmogorov's famous four-fifths law,

\begin{align}
S_3\left(r\right) = -\frac{4}{5}\langle\epsilon\rangle r.
\label{eq:45}
\end{align}

\begin{figure}
	\centering
	\includegraphics[width=0.5\textwidth]{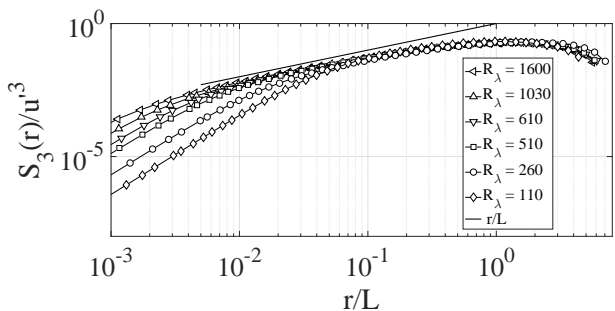}
	\caption{Third-order structure functions measured in the Variable Density Turbulence Tunnel. Here, $u'$ is the fluctuating velocity and $L$ the integral length scale. The straight black line is equal to $r/L$, the scaling predicted by equation \ref{eq:45}.}
	\label{fig:s3}
\end{figure}

The variation of the third-oder structure function with the Reynolds number can be seen in figure \ref{fig:s3}. With the Reynolds number approaching higher values, an inertial range emerges and the third-order structure function appears to fullfill Kolmogorov's famous four-fifths law

Further assuming self-similarity of the turbulent flow at different scales, one can generalize the result to structure functions of arbitrary orders $n$, with unknown constants $C_n$, 

\begin{align}
\label{eq:k41caling}
S_n\left(r\right) = C_n\left(\varepsilon r\right)^{n/3}.
\end{align}

\begin{figure}
	\includegraphics[width=0.5\textwidth]{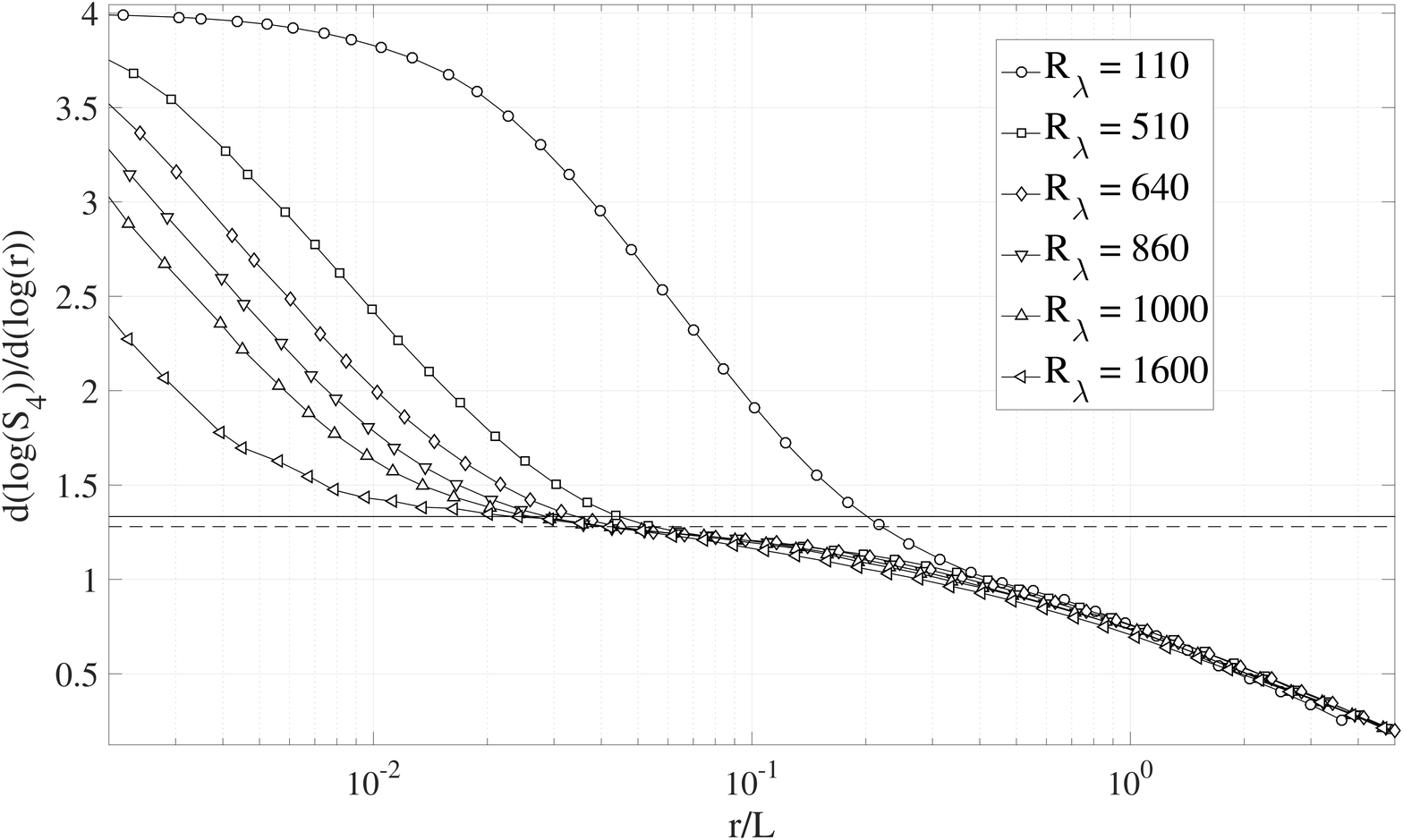}
	\caption{Logarithmic derivative of the fourth-order structure function with respect to the separation. Even at the highest Reynolds numbers measured, there seems to be no approach to a horizontal line that would correspond to power-law scaling. The solid horizontal line is the K41 prediction, the dashed line the prediction of the model by \citet{She1994}.}
	\label{fig:s4r}
\end{figure}

In this framework, the scaling exponents of the structure functions $\zeta_n = n/3$ are simply a linear function of the order. For real turbulent flows, the assumption of self-similarity does not hold and numerous refined models have been proposed to describe structure functions in the inertial range. \citet{Kolmogorov1962} allowed for intermittency of the energy dissipation rate following a comment by \citet{Landau1959} resulting in the K62 framework. The $\beta$-model by \citet{Frisch1978} applied concepts of fractal dimensions to turbulent flows, being refined by \citet{Benzi1984} in the random-$\beta$-model. The $p$-model by \citet{Meneveau1987a} and \citet{Meneveau1987b} allows for unequal distribution of energy in the eddy-breaking process of the energy cascade. The $\gamma$-model by \citet{Andrews1989} proposed a $y$-distribution of the energy dissipation, while \citet{Kida1991}
assumed a log-stable distribution. \citet{She1994} built a parameter-free model based upon a hierarchy of dissipation moments. This was extended by \citet{Dubrulle1994}. All of these models have in common power-law scaling of the structure functions in the inertial range, with varying predictions for the scaling exponents as a nonlinear function of the order.
Assuming the validity of these predictions one should be able to extract scaling exponents from the data by computing logarithmic derivatives $d\log(S_n)/d\log(r)=\zeta_n$. A power-law behavior here would correspond to a horizontal line in a graph of $\mathrm{d}\log S_n/d\log r$ against $r$. All deviations from this line are connected to deviations from power-law behavior (e.g. \cite{Sreenivasan1998}). Such a graph of our data is shown in figure \ref{fig:s4r}. Despite the large Reynolds numbers however, there seems to be no obvious approach to true scaling.

This lack of observable scaling is a common feature in experimental turbulent flows and led to an extraction method developed by \citet{Benzi1993} called Extended Self-Similarity (ESS). Rather than scaling with respect to the separation $r$, it is proposed that structure functions scale with respect to each other, namely $S_n\propto S_m^{\zeta_{m,n}}$ with the relative scaling exponents $\zeta_{m,n}$. Note that different definitions of $S_m$ can influence the resulting scaling exponents \citep{vandeWater1995}.

%\section{The Experiment}

We performed the experiments in the Variable Density Turbulence Tunnel (the VDTT) 
at the Max Planck Institute for Dynamics and Self-Organization \citep{Bodenschatz2014}.  
The wind tunnel was a pressurizable closed-circuit wind tunnel, 
in which either air or sulfur hexafluoride (SF$_6$) circulated.  
By changing the pressure of the gas, we could adjust its kinematic viscosity, $\nu$, 
and thus the Reynolds number, $R_\lambda$, 
without changing the geometrical boundary conditions or the mean speed of the flow.  
We set the pressure to values between %0.5 %
1 and 15 bar, 
and so prepared viscosities between $1.55\cdot10^{-5} m^2/s$ and $1.42\cdot 10^{-7}$ $m^2/s$.  

To produce turbulence, we used a bi-planar grid of square grid bars of mesh spacing of 18\,cm 
that blocked about 40\% of the cross section of the tunnel.  
The mean speed, $U$, of the flow was kept constant at 4.2\,m/s.  
The temperature of the gas was stable to within 0.2\,K over arbitrary times and was set to fixed values between 22.0$^{\circ}$C and 23.4$^{\circ}$C.

The velocity of the gas was measured at either 7.1\,m or 8.3\,m downstream of the grid 
with the NSTAP probes developed at Princeton university \citep{Bailey2010,Vallikivi2011}.  
These were micro-fabricated hot-wire probes, 
manufactured in two different ways such that they had 
either 30 or 60 micron long sensor elements.  
In addition to the data acquired with the NSTAPs, 
we also acquired data with larger hot-wire probes of traditional construction 
produced by Dantec Dynamics, 0.45 or 1.25\,mm long.  
The data acquired with the Dantec probes are not shown in this paper, but they gave results consistent with those from the NSTAP and the smaller Dantec probes.
The longer datasets presented in this paper were acquired in conjunction with those shown 
in our paper on the 
decay of turbulence \citep{Sinhuber2015}, 
but the present data are much longer 
as needed for the different purpose of this paper - providing sufficient statistics to uncover details of the inertial-range statistics.

The quantity measured by the hot-wire probes 
is the velocity time series $v \left( t \right)$ 
observed at a single position, 
low-pass filtered at either 30\,kHz or 100\,kHz, 
and sampled at either 60\,kHz or 200\,kHz, respectively.  
The turbulence intensity $u'$ of the tunnel, 
was between 1.6\% and 3.6\% of the mean speed $U$, 
which is small enough safely to invoke Taylor's hypothesis \citep{Taylor1938} 
in order to convert the functions of time, $v \left( t \right)$, 
to functions of space, so that $v \left( tU \right) = v \left( x \right)$.  

We divide the various data into two categories, 
which we call datasets A and B. Each dataset is comprised of measurements made at different Reynolds numbers.  
What separates the datasets is the length of the velocity records.  
Dataset A consisted of 14 measurements, which were each between 6 and 9 hours long, 
or $\mathcal{O}(10^6)$ integral scales long.  
Dataset B consisted of 4 measurements, which were each between 2 and 3 days long, 
or $\mathcal{O}(10^7)$ integral scales long. Take note that the amount of data obtained in these datasets exceeds any comparable experiment by an order of magnitude, making it possible to investigate fine details in the inertial-range structure of turbulence.
The integral length scale for all flows was about 0.1\,m.  
Altogether, the data sets span Taylor Reynolds numbers between 110 and 1600.

%\section{Results}

The aforementioned models considered statistics in the inertial range where, as in the classical cascade model, neither dissipation nor energy injection play a role. However, in real turbulence, there is no sharp distinctive scale between the ranges where the statistics change. Even well above the Kolmogorov length, dissipation influences the statistics, leading to the so-called near-dissipation range \citep{Frisch1991}. A successful model describing the effects of dissipation in the near-dissipation regime \citep{Frisch1985} has been developed by \citet{Meneveau1996}. He noted that an order-dependent viscous cutoff-scale leads to deviations from scaling behavior in the near-dissipation range that grow larger as the Reynolds number increases. There are numerous alternative models for the near-dissipation range for example by \citet{She1991}, \citet{Biferale1993} and \citet{Chevillard2005}. Additional detailed investigations of the transition between the dissipative range and the inertial range have been conducted by e.g.  \cite{Falkovich1994} and \cite{Donzis2010} in terms of the bottleneck effect of the energy spectrum. It has been further shown that ringing in the energy spectrum also affects the transition between different ranges of structure functions and their Fourier-space representations (\cite{Lohse1995}, \cite{Lohse1996} and \cite{Dobler2003}). 
We do not discriminate between the models and chose to compare our data with Meneveau's multifractal model for convenience. 
In the multifractal model, the structure functions are functions of both $r/\eta$ and $r/L$,

\begin{align}
S_n = f_n\left(r/\eta\right)\left(\frac{r}{L}\right)^{\zeta_n}.
\end{align} 

Using the the predictions of the $p$-model by \citet{Meneveau1987a} and \citet{Meneveau1987b}, \citet{Meneveau1996} computes the functional form of  $f_n\left(r/\eta\right)$. The scaling exponents of the $p$-model are nonlinear functions of n \citep{Meneveau1987b},

\begin{align}
\zeta_n = 1-log_2\left(p^{n/3}+\left(1-p\right)^{n/3}\right).
\end{align}

Allowing for an order-dependent viscous cutoff scale, \citet{Meneveau1996} computes the functional form for the structure functions as

\begin{align}
f_n\left(r/\eta\right) =\left(1+\left(c_a\frac{r}{\eta}\left(c_b\cdot R_\lambda^{3/2}\right)^{\frac{1-\alpha_p}{3+\alpha_p}}\right)^{-2}\right)^{\left(\zeta_n-n\right)/2}
\end{align}

with $c_a= 0.1\cdot15^{-0.75}$, $c_b = 1/13$ and

\begin{align}
\alpha_p = -\frac{p^{n/3}\log_2p+(1-p)^{n/3}\log_2(1-p)}{p^{n/3}+(1-p)^{n/3}}.
\end{align}

As in the original publication, we set the free parameter $p$ to 0.7. $p$ governs how unequally energy is distributed among eddies breaking during the turbulent cascade.

\begin{figure}
	\includegraphics[width=0.5\textwidth]{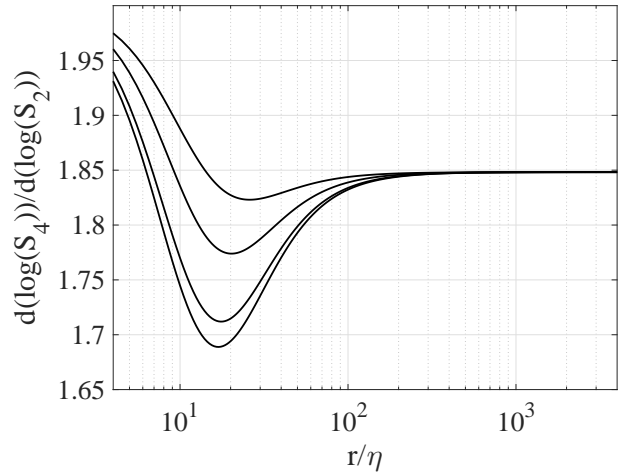}
	\caption{Prediction for the shape of the logarithmic derivative of the fourth-order structure function with respect to the second-order structure function as given by the multifractal model \citep{Meneveau1996} using the p-model \citep{Meneveau1987a} for the scaling exponents and turbulence parameters from several VDTT-datasets for $R_\lambda$ between 110 and 1600. The model predicts a single oscillation with a minimum at about $20\eta$.}
	\label{fig:s4s2model}
\end{figure}

Figure \ref{fig:s4s2model} shows ESS-plots from the multifractal model. We present the logarithmic derivative of the fourth-order structure function by the second-order structure functions using the scaling exponent prediction of the $p$-model to compute the model structure functions. Any other choice of structure functions or scaling exponent model would have given qualitatively equivalent results. The free turbulence parameters in this plot, $\eta$, $L$ and $R_\lambda$ are extracted from datasets A and B for 5 datasets between $R_\lambda = 110$ (top) and $R_\lambda = 1600$ (bottom). In the multifractal model there is a significant minimum in the near-dissipation range at around $20\eta$ as a result of the order-dependent cutoff scale \citep{Frisch1991}. For $r\gg20\eta$ the logarithmic derivative approaches its expected inertial-range limit of $\zeta_{4,2}$. Within the multifractal model, there is strict power-law scaling in the structure functions as long as $r\gg20\eta$.

\begin{figure}
	\includegraphics[width=0.5\textwidth]{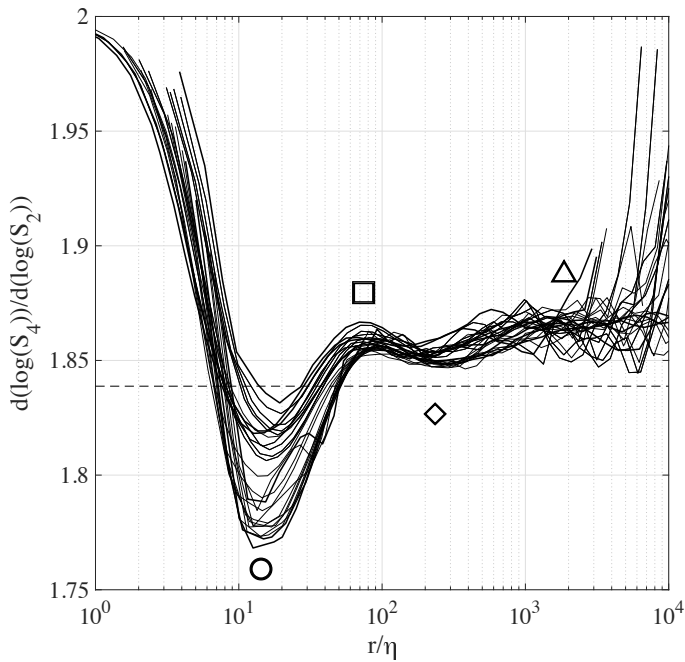}
	\caption{Logarithmic derivative of the fourth-order structure with respect to the second-order structure function, obtained by NSTAP probes. The thick lines correspond to dataset B, thin lines to dataset A. The data span Taylor-scale Reynolds numbers from 110 to 1600. For comparison, the prediction of the model by \citet{She1994} is indicated by a dashed line. Note that the K41 prediction is 2. The near-dissipation range deviations are in agreement with the predictions of the multifractal model by \citet{Meneveau1996}; there is an overshoot at about $20\eta$ ($\circ$). However, more structural details are observable in the inertial range, being uncovered by the large amount of statistics. These additional oscillatory features are marked by $\square$, $\diamond$ and $\triangle$.}
	\label{fig:s4s2}
\end{figure}

It has since been a commonly accepted procedure to extract scaling exponents from ESS plots by averaging over scales that are reasonably far away from the Kolmogorov scale, in accordance with the multifractal predictions. However, the high-Reynolds number, extremely long datasets from the VDTT imply the need for a more refined interpretation. Figure \ref{fig:s4s2} shows the logarithmic derivative of the fourth-order structure function with respect to the second-order structure function for the whole range of Reynolds numbers between $R_\lambda = 100$ and $R_\lambda = 1600$ in datasets A and B. In the near-dissipation range, between $10\eta$ and $30\eta$ the data are in good qualitative agreement with the predictions of the multifractal model as seen in \ref{fig:s4s2model}. For instance, with increasing Reynolds number, the minimum in the near-dissipation range ($\circ$) grows significantly in depth. At smaller scales, the NSTAP data starts to get distorted by electric noise from the CTA measurement system and temporal and spatial filtering effects begin to influence the probe response. Nonetheless, our experiments uncover features not reported in the literature. While the multifractal model predicts a monotonic approach from the near-dissipation range minimum toward the value of the ratio of the scaling exponents, the VDTT data show an overshoot ($\square$) at around $70\eta$, which is approximately independent of Renyolds number. In contrast to the expectations, the data from the VDTT do not approach a constant value at higher $r/\eta$. Instead, an only weakly Reynolds-number dependent and persistent substructure in the inertial range becomes visible.

\begin{figure}
	\includegraphics[width=0.5\textwidth]{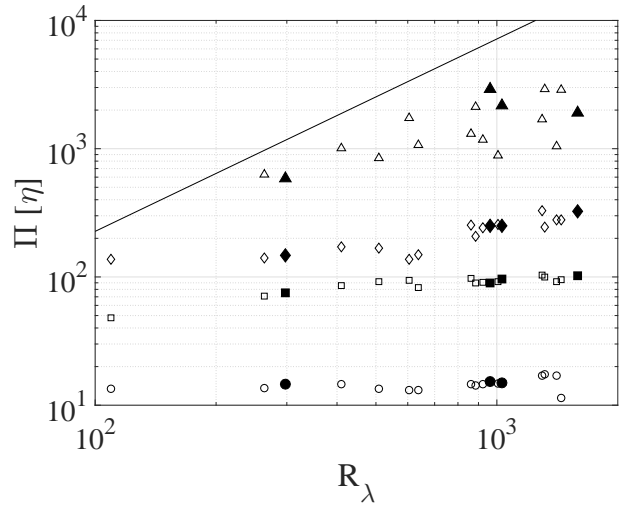}
	\caption{The positions $\Pi$ of the extrema in the logarithmic derivatives of figure \ref{fig:s4s2} in terms of the Kolmogorov scale $\eta$ as a function of the Reynolds number $R_\lambda$. The solid black line is $0.5\cdot L/\eta$, predicted by our experiments. Circles represent the location of the near-dissipation range minumum, squares the first maximum, diamonds the second minimum and triangles the second maximum. Open symbols correspond to dataset A, filled symbols to dataset B. Once the Reynolds number is high enough, the features denoted by circles, squares and diamonds and triangles appear to be only weakly dependent on the Reynolds number.}
	\label{fig:extrema}
\end{figure}

Figure \ref{fig:extrema} shows the positions, $\Pi$, of the substructures in figure \ref{fig:s4s2} in terms of the Kolmogorov scale and as function of the Reynolds number. These positions have been obtained using a windowed parabolic fit to the data up to $r=2L$, though other methods such as 4-peak Gaussian fits yield comparable results. In addition to the overshoot ($\square$) at about $70\eta$, two more extrema become apparent in figure \ref{fig:s4s2} ($\diamond$ and $\triangle$) before the data succumbs to noise at larger $r/\eta$. These structures in the inertial range only become visible with the long datasets from the VDTT at high Reynolds number, at either lower Reynolds numbers; or significantly shorter datasets, these extrema are lost within the noise. These new inertial-range structures in the structure functions are associated with the bottleneck phenomenon in the energy spectrum \cite{Falkovich1994} and can be embedded in refinements of existing inertial-range models. The features are reminiscent of the lacunarity proposed by \citet{Smith1986}.

%\section{Discussion}

These findings have significant implications for the interpretation of the statistical behavior of turbulent flows. The persistent structures in the ESS-plots are in contradiction with the prediction of a power-law behavior in the inertial range, and imply that there is an oscillatory component in the structure functions themselves, albeit small, as an order-dependent cutoff scale as in the multifractal model is by construction unable to include multiple overshoots. A consequence is that with different choices of averaging intervals the same data yield different scaling exponents. The non-power law behavior demands refined models of the structure functions. The maxima and minima we find in the inertial range are not strongly dependent on the Reynolds number and are anchored to the small scales $\eta$. This unexpected behavior shows that far away from the dissipation range, far even from the near-dissipation range, dissipative effects still qualitatively affect the statistics of the flow and it implies meso-scale organization of small-scale structures. These findings might be associated with more complex cascade processes hinted in \cite{Meneveau1987b}. It is not clear from our data how far up from the dissipation range in scale the oscillations persist. Future investigations should measure at even higher Reynolds number to achieve a higher ratio of $L/\eta$ and to uncover more features of the inertial-range statistics. With an active grid installed, the VDTT is able to achieve Reynolds number significantly higher than the classic grid data presented here and will be such a further step.

\section{Acknowledgments}
We thank A. Renner,
A. Kopp, A. Kubitzek, H. Nobach, U. Schminke, and the
machinists at the MPI-DS who helped to build and maintain the VDTT.
The NSTAPs were developed at Princeton University by M. Vallikivi, M. Hultmark and A. J. Smits  who graciously provided them and helped to make them work with the VDTT equipment. We are thankful to F. K\"{o}hler and L. Hillmann
who assisted in acquiring the data as well as to M. Wilczek and W. van de Water for fruitful discussions.
The data analyzed here was taken at the Max-Planck-Institute for Dynamics and Self-Organization during the doctoral thesis work of Michael Sinhuber.

\bibliographystyle{apsrev}

%\bibliography{literature.bib}

\begin{thebibliography}{36}
	\expandafter\ifx\csname natexlab\endcsname\relax\def\natexlab#1{#1}\fi
	
	\bibitem[Andrews {\em et~al.\/}(1989)Andrews, Phillips, Shivamoggi, Beck \&
	Joshi]{Andrews1989}
	{\sc Andrews, L.~C., Phillips, R.~L., Shivamoggi, B.~K., Beck, J.~K. \& Joshi,
		M.~L.} 1989 A statistical theory for the distribution of energy dissipation
	in intermittent turbulence. {\em Phys. Fluids A\/} {\bf 1}, 999--1006.
	
	\bibitem[Bailey {\em et~al.\/}(2010)Bailey, Kunkel, Hultmark, Vallikivi, Hill,
	Meyer, Tsay, Arnold \& Smits]{Bailey2010}
	{\sc Bailey, S. C.~C., Kunkel, G.~J., Hultmark, M., Vallikivi, M., Hill, J.~P.,
		Meyer, K.~A., Tsay, C., Arnold, C.~B. \& Smits, A.~J.} 2010 Turbulence
	measurements using a nanoscale thermal anemometry probe. {\em J. Fluid
		Mech.\/} {\bf 663}, 160--179.
	
	\bibitem[Benzi {\em et~al.\/}(1993)Benzi, Ciliberto, Tripiccione, Baudet,
	Massaioli \& Succi]{Benzi1993}
	{\sc Benzi, R., Ciliberto, S., Tripiccione, R., Baudet, C., Massaioli, F. \&
		Succi, S.} 1993 Extended self-similarity in turbulent flows. {\em Phys. Rev.
		E\/} {\bf 48}, R29--R32.
	
	\bibitem[Benzi {\em et~al.\/}(1984)Benzi, Paladin, Parisi \&
	Vulpiani]{Benzi1984}
	{\sc Benzi, R., Paladin, G., Parisi, G. \& Vulpiani, A.} 1984 On the
	multifractal nature of fully developed turbulence and chaotic systems. {\em
		J. Phys. A: Math. Gen.\/} {\bf 17}, 3521--3531.
	
	\bibitem[Biferale(1993)]{Biferale1993}
	{\sc Biferale, L.} 1993 Probability distribution functions in turbulent flows
	and shell models. {\em Phys. Fluids A\/} {\bf 5}, 428--435.
	
	\bibitem[Bodenschatz {\em et~al.\/}(2014)Bodenschatz, Bewley, Nobach, Sinhuber
	\& Xu]{Bodenschatz2014}
	{\sc Bodenschatz, E., Bewley, G.~P., Nobach, H., Sinhuber, M. \& Xu, H.} 2014
	Variable density turbulence tunnel facility. {\em Rev. Sci. Instrum.\/} {\bf
		85}, 093908.
	
	\bibitem[Chevillard {\em et~al.\/}(2005)Chevillard, Castaing \&
	L\'{e}v\^{e}que]{Chevillard2005}
	{\sc Chevillard, L., Castaing, B. \& L\'{e}v\^{e}que, E.} 2005 On the rapid
	increase of intermittency in the near-dissipation range of fully developed
	turbulence. {\em Eur. Phys. J. B\/} {\bf 45}, 561--567.
	
	\bibitem[Dobler {\em et~al.\/}(2003)Dobler, Haugen, Yousef \&
	Brandenburg]{Dobler2003}
	{\sc Dobler, W., Haugen, N. E.~L., Yousef, T.~A. \& Brandenburg, A.} 2003
	Bottleneck effect in three-dimensional turbulence simulations. {\em Phys.
		Rev. E\/} {\bf 68}, 026304--026311.
	
	\bibitem[Donzis \& Sreenivasan(2010)]{Donzis2010}
	{\sc Donzis, D.~A. \& Sreenivasan, K.~R.} 2010 The bottleneck effect and the
	kolmogorov constant in isotropic turbulence. {\em J. Fluid Mech.\/} {\bf
		657}, 171--188.
	
	\bibitem[Dubrulle(1994)]{Dubrulle1994}
	{\sc Dubrulle, B.} 1994 Intermittency in fully developed turbulence:
	Log-poisson statistics and generalized scale covariance. {\em Phys. Rev.
		Lett.\/} {\bf 73}, 959--962.
	
	\bibitem[Falkovich(1994)]{Falkovich1994}
	{\sc Falkovich, G.} 1994 Bottleneck phenomenon in developed turbulence. {\em
		Phys. Fluids\/} {\bf 6}, 1411--1414.
	
	\bibitem[Frisch \& Parisi(1985)]{Frisch1985}
	{\sc Frisch, U. \& Parisi, G.} 1985 Fully developed turbulence and
	intermittency. In {\em Turbulence and predictability in geophysical fluid
		dynamics and climate dynamics\/}, , vol.~88, pp. 71--88. Elsevier Science
	Ltd.
	
	\bibitem[Frisch {\em et~al.\/}(1978)Frisch, Sulem \& Nelkin]{Frisch1978}
	{\sc Frisch, U., Sulem, P.-L. \& Nelkin, M.} 1978 A simple dynamical model of
	intermittent fully developed turbulence. {\em J. Fluid Mech.\/} {\bf 87},
	719--736.
	
	\bibitem[Frisch \& Vergassola(1991)]{Frisch1991}
	{\sc Frisch, U. \& Vergassola} 1991 A prediction of the multifractal model: the
	intermediate dissipation range. {\em Europhys. Lett.\/} {\bf 14}~(5),
	439--444.
	
	\bibitem[de~K\'arm\'an \& Howarth(1938)]{Karman1938}
	{\sc de~K\'arm\'an, T. \& Howarth, L.} 1938 On the statistical theory of
	isotropic turbulence. {\em Proc. Roy. Soc. Lond. A\/} {\bf 164}, 192--215.
	
	\bibitem[Kida(1991)]{Kida1991}
	{\sc Kida, S.} 1991 Log-stable distribution and intermittency of turbulence.
	{\em J. Phys. Soc. JPN.\/} {\bf 60}, 5--8.
	
	\bibitem[Kolmogorov(1941{\natexlab{{\em a\/}}})]{Kolmogorov1941b}
	{\sc Kolmogorov, A.~N.} 1941{\natexlab{{\em a\/}}} Dissipation of energy in
	locally isotropic turbulence. {\em Dokl. Akad. Nauk SSSR\/} {\bf 32}, 16--18.
	
	\bibitem[Kolmogorov(1941{\natexlab{{\em b\/}}})]{Kolmogorov1941a}
	{\sc Kolmogorov, A.~N.} 1941{\natexlab{{\em b\/}}} The local structure of
	turbulence in incompressible viscous fluid for very large reynolds numbers.
	{\em Dokl. Akad. Nauk SSSR\/} {\bf 30}, 299--303.
	
	\bibitem[Kolmogorov(1962)]{Kolmogorov1962}
	{\sc Kolmogorov, A.~N.} 1962 A refinement of previous hypotheses concerning the
	local structure of turbulence in a viscous incompressible fluid at high
	reynolds number. {\em J. Fluid Mech.\/} {\bf 13}, 82--85.
	
	\bibitem[Landau \& Lifschitz(1959)]{Landau1959}
	{\sc Landau, L.~D. \& Lifschitz, E.~M.} 1959 {\em Fluid Mechanics\/}. Pergamon
	Press, translated from Russian by J. B. Sykes and W. H. Reid.
	
	\bibitem[Lohse \& M\"uller-Groeling(1995)]{Lohse1995}
	{\sc Lohse, D. \& M\"uller-Groeling, A.} 1995 Bottleneck effects in turbulence:
	Scaling phenomena in $r$ versus $p$ space. {\em Phys. Rev. Lett.\/} {\bf
		74}~(10), 1747--1750.
	
	\bibitem[Lohse \& M\"uller-Groeling(1996)]{Lohse1996}
	{\sc Lohse, D. \& M\"uller-Groeling, A.} 1996 Anisotropy and scaling
	corrections in turbulence. {\em Phys. Rev. E\/} {\bf 54}~(1), 395--405.
	
	\bibitem[Meneveau(1996)]{Meneveau1996}
	{\sc Meneveau, C.} 1996 Transition between viscous and inertial-range scaling
	of turbulence structure functions. {\em Phys. Rev. E\/} {\bf 54}~(4),
	3657--3663.
	
	\bibitem[Meneveau \& Sreenivasan(1987{\natexlab{{\em a\/}}})]{Meneveau1987a}
	{\sc Meneveau, C. \& Sreenivasan, K.~R.} 1987{\natexlab{{\em a\/}}} The
	multifractal spectrum of the dissipation field in turbulent flows. {\em Nucl.
		Phys. B\/} {\bf 2}, 49--76.
	
	\bibitem[Meneveau \& Sreenivasan(1987{\natexlab{{\em b\/}}})]{Meneveau1987b}
	{\sc Meneveau, C. \& Sreenivasan, K.~R.} 1987{\natexlab{{\em b\/}}} Simple
	multifractal cascade model for fully developed turbulence. {\em Phys. Rev.
		Lett.\/} {\bf 59}, 1424--1427.
	
	\bibitem[Navier(1827)]{Navier1827}
	{\sc Navier, C. L.~M.} 1827 Sur les lois du mouvement des fluids. {\em Comptes
		Rendus des Seances de l'Academie des Sciences\/} {\bf 6}, 389--440.
	
	\bibitem[Richardson(1922)]{Richardson1922}
	{\sc Richardson, L.~F.} 1922 {\em Weather prediction by numerical process\/}.
	Cambridge University Press.
	
	\bibitem[She(1991)]{She1991}
	{\sc She, Z.-S.} 1991 Physical model of intermittency in turbulence:
	Near-dissipation-range non-gaussian statistics. {\em Phys. Rev. Lett.\/} {\bf
		66}~(5), 600--603.
	
	\bibitem[She \& L\'{e}v\^{e}que(1994)]{She1994}
	{\sc She, Z.-S. \& L\'{e}v\^{e}que, E.} 1994 Universal scaling laws in fully
	developed turbulence. {\em Phys. Rev. Lett.\/} {\bf 72}, 336--339.
	
	\bibitem[Sinhuber {\em et~al.\/}(2015)Sinhuber, Bodenschatz \&
	Bewley]{Sinhuber2015}
	{\sc Sinhuber, M., Bodenschatz, E. \& Bewley, G.~P.} 2015 Decay of turbulence
	at high reynolds numbers. {\em Phys. Rev. Lett.\/} {\bf 114}, 034501.
	
	\bibitem[Smith {\em et~al.\/}(1986)Smith, Fournier \& Spiegel]{Smith1986}
	{\sc Smith, L.~A., Fournier, J.-D. \& Spiegel, E.~A.} 1986 Lacunarity and
	intermittency in fluid turbulence. {\em Phys. Lett.\/} {\bf 114A}~(8,9),
	465--468.
	
	\bibitem[Sreenivasan \& Dhruva(1998)]{Sreenivasan1998}
	{\sc Sreenivasan, K.~R. \& Dhruva, B.} 1998 Is there scaling in
	high-reynolds-number turbulence. {\em Prog. Theor. Phys. Supp.\/} {\bf 130},
	103--120.
	
	\bibitem[Stokes(1845)]{Stokes1845}
	{\sc Stokes, G.~G.} 1845 On the theories of the internal friction of fluids in
	motion, and of the equilibrium and motion of elastic solids. {\em Tran. Camb.
		Phil. Soc.\/} {\bf 8}, 287--305.
	
	\bibitem[Taylor(1938)]{Taylor1938}
	{\sc Taylor, G.~I.} 1938 The spectrum of turbulence. {\em Proc. Roy. Soc. Lond.
		A\/} {\bf 164}, 476--490.
	
	\bibitem[Vallikivi {\em et~al.\/}(2011)Vallikivi, Hultmark, Bailey \&
	Smits]{Vallikivi2011}
	{\sc Vallikivi, M., Hultmark, M., Bailey, S. \& Smits, A.} 2011 Turbulence
	measurements in pipe flow using a nano-scale thermal anemometry probe. {\em
		Exp. Fluids\/} {\bf 51}, 1521--1527.
	
	\bibitem[van~de Water \& Herweijer(1995)]{vandeWater1995}
	{\sc van~de Water, W. \& Herweijer, J.} 1995 Comment on "extended
	self-similarity in turbulent flows". {\em Phys. Rev. E\/} {\bf 51}~(3),
	2669--2671.
	
\end{thebibliography}

\end{document}